\documentclass[lettersize,journal]{IEEEtran}
\usepackage{amsmath,amsfonts,bm,bbm}
\usepackage{algorithmic}
\usepackage{algorithm}
\usepackage{array}
\usepackage[caption=false,font=normalsize,labelfont=sf,textfont=sf]{subfig}
\usepackage{textcomp}
\usepackage{stfloats}
\usepackage{url}
\usepackage{verbatim}
\usepackage{graphicx}
\usepackage{cite}
\usepackage{pifont}
\usepackage{siunitx}
\usepackage{color}
\usepackage[colorlinks,linkcolor=blue]{hyperref}
\newcommand{\diag}{{\mathrm{diag}}}
\newcommand{\hbx}{\hat{\mathbf{x}}}
\newcommand{\tbx}{\tilde{\mathbf{x}}}
\newcommand{\hbP}{\hat{\mathbf{P}}}
\newcommand{\iN}{\check{\cN}}
\newcommand{\oN}{\hat{\cN}}

\newcommand{\tp}{\mathsf{T}}
\newcommand{\gam}{\gamma}

\newcommand{\bsl}{\boldsymbol}
\newcommand{\ul}{\underline}
\newcommand{\ol}{\overline}

\def\bA{\mathbf{A}}

\def\bB{\mathbf{B}}

\def\bC{\mathbf{C}}

\def\bD{\mathbf{D}}

\def\bF{\mathbf{F}}

\def\bh{\mathbf{h}}
\def\bH{\mathbf{H}}

\def\bI{\mathbf{I}}

\def\bK{\mathbf{K}}

\def\bM{\mathbf{M}}
\def\bn{\mathbf{n}}

\def\bP{\mathbf{P}}
\def\bq{\mathbf{q}}
\def\bQ{\mathbf{Q}}

\def\bR{\mathbf{R}}

\def\bS{\mathbf{S}}

\def\bv{\mathbf{v}}
\def\bV{\mathbf{V}}
\def\bw{\mathbf{w}}

\def\bx{\mathbf{x}}

\def\by{\mathbf{y}}

\def\bz{\mathbf{z}}

\def\cH{\mathcal{H}}

\def\cN{\mathcal{N}}

\def\cR{\mathcal{R}}

\def\cV{\mathcal{V}}

\def\cX{\mathcal{X}}

\newtheorem{theorem}{Theorem}
\newtheorem{lemma}{Lemma}

\newtheorem{assumption}{Assumption}

\allowdisplaybreaks[3]

\begin{document}

\title{Distributed Target Tracking with Fading Channels over Underwater Wireless Sensor Networks}

\author{Miaoyi Tang, Meiqin Liu, \IEEEmembership{Senior Member, IEEE}, Senlin Zhang, \IEEEmembership{Member, IEEE}, Ronghao Zheng, \IEEEmembership{Member, IEEE},  and Shanling Dong, \IEEEmembership{Member, IEEE}
	\thanks{This work was supported by the National Natural Science Foundation of China under Grants 62173299 and U1909206, the Zhejiang Provincial Natural Science Foundation of China under Grant LZ23F030006, the Joint Fund of Ministry of Education for Pre-research of Equipment under Grant 8091B022147, and the Fundamental Research Funds for the Central Universities under Grant xtr072022001. (Corresponding author: Meiqin Liu.)}
	\thanks{Miaoyi Tang, Senlin Zhang, Ronghao Zheng, and Shanling Dong are with the National Key Laboratory of Industrial Control Technology and the College of Electrical Engineering, Zhejiang University, Hangzhou 310027, China (e-mail:
		12110073@zju.edu.cn, slzhang@zju.edu.cn, rzheng@zju.edu.cn, shanlingdong28@zju.edu.cn).}
	\thanks{Meiqin Liu is with the Institute of Artificial Intelligence and Robotics,
		Xi’an Jiaotong University, Xi’an 710049, China and also with the College of Electrical Engineering, Zhejiang University, Hangzhou
		310027, China (e-mail:liumeiqin@zju.edu.cn).}
}

\markboth{}%
{Shell \MakeLowercase{\textit{et al.}}: A Sample Article Using IEEEtran.cls for IEEE Journals}

\IEEEpubid{}

\maketitle

\begin{abstract}
This paper investigates the problem of distributed target tracking via underwater wireless sensor networks (UWSNs) with fading channels. The degradation of signal quality due to wireless channel fading can significantly impact network reliability and subsequently reduce the tracking accuracy. To address this issue, we propose a modified distributed unscented Kalman filter (DUKF) named DUKF-Fc, which takes into account the effects of measurement fluctuation and transmission failure induced by channel fading. The channel estimation error is also considered when designing the estimator and a sufficient condition is established to ensure the stochastic boundedness of the estimation error. The proposed filtering scheme is versatile and possesses wide applicability to numerous real-world scenarios, e.g., tracking a maneuvering underwater target with acoustic sensors. Simulation results demonstrate the effectiveness of the proposed filtering algorithm. In addition, considering the constraints of network energy resources, the issue of investigating a trade-off between tracking performance and energy consumption is discussed accordingly.
\end{abstract}

\begin{IEEEkeywords}
distributed state estimation, unscented Kalman filtering, channel fading, underwater wireless sensor networks, stochastic stability.
\end{IEEEkeywords}

\section{Introduction}

\IEEEpubidadjcol

\IEEEPARstart{W}{ireless} sensor networks (WSNs) have attracted increasing attention in recent years, driven by advancements in communication, sensing, and computing technologies. Their potential for various applications, such as environmental monitoring, surveillance, and industrial automation, stems from the inherent advantages of wireless communication, including flexibility, adaptability, and so forth [1]-[3].

However, challenges arise in terms of communication stability, as wireless channels are more susceptible to environmental factors compared to wired communication [4]-[7]. In general, wireless channels exhibit time-variability due to factors such as moving objects, changing environmental conditions, and varying transmitter/receiver positions [8]. As a result, the use of wireless channels can lead to slow or fast fading, attributable to various effects, including variations in multipath propagation, shadowing, and other factors [9]-[11]. These effects can significantly degrade network reliability and the accuracy of data collected by sensors cannot be ignored when designing algorithms for WSNs.

Tracking maneuvering target is an important area of interest in the field of WSNs applications. Among target tracking techniques, the Kalman filter is a widely used tool for state estimation in dynamic systems, and the impact of channel fading on the filter stability has been extensively studied [12]-[19]. In [12], the issue of adapting Kalman filtering in a sensor network with noisy fading wireless channels was discussed. The expected error covariance for scalar measurements was shown to converge to a steady-state value and was bounded with respect to the fading process.  Considering the relationship between packet loss probabilities and transmission power level, a time-varying Kalman filter and a predictive control algorithm were employed in [13] to optimize the trade-off between the energy expended by the sensors and the accuracy of state estimation. Based on the above works, [14] investigated the stochastic stability of Kalman filtering with wireless fading channels and formulated stabilizing power control policies that minimize the total sensor power, whereas [15]-[17] extended these approaches to the case of correlated fading channels. Besides, the effects of fading-induced measurement fluctuation on the filter stability were discussed in [18], [19]. The latter work examined the issue of nonlinear state estimation and proposed a modified unscented Kalman filter (UKF) in the presence of fading measurements and transmission failure.

\IEEEpubidadjcol

Distributed state estimation (DSE) techniques have emerged as promising solutions for target tracking in sensor networks [20]. Following the principle of distributed computation, these techniques offer several advantages over centralized state estimation, including improved network robustness, reduced communication overhead, and enhanced resilience to node failures [21], [22], which collectively contribute to stabilizing the tracking performance under non-ideal conditions. For example, the works [23]-[25] focused on the issue of DSE subject to random communication failures and addressed the problem of stability by using consensus-based approaches. Based on [18], a recent work [26] considered both measurements fading and random link failures and built the DSE algorithm consisting of a structural data fusion stage and a signal data fusion stage. As for nonlinear systems, the authors in [27] proposed an event-triggered distributed extended Kalman filter (DEKF), where a variance-constrained approach was adopted to derive the upper bound for the estimation error covariance by considering the errors originated from fading measurements. Further in [28], a distributed unscented Kalman filter (DUKF) algorithm with intermittent measurements was proposed based on the principle of information filter (IF), where the covariance intersection (CI) strategy was introduced to enhance the estimation performance. 

Of note, the above works are founded on the assumption of accurate channel states knowledge, i.e., the fadings of signal can be precisely measured. However, in real-time applications, the estimation of time-varying channel states can rarely be error-free [29]-[32], and the corresponding estimation error may result in the instability of the filter process. To the best of our knowledge, relatively little research has been carried out on this issue in the field of DSE, especially for nonlinear systems. Motivated by this, we propose a modified distributed unscented Kalman filtering (DUKF) algorithm called DUKF-Fc in this paper, which enables the wireless sensor nodes to perform nonlinear state estimation while communicating over fading channels. In the algorithm design, we take into account the channel estimation error as well as the effects induced by channel fading including transmission failure and measurement fluctuation. Further, we conduct a performance analysis of DUKF-Fc and prove that the estimation error is bounded in mean square.

In addition to the theoretical analysis, the proposed methodology is applied in the context of target tracking using underwater wireless sensor networks (UWSNs) equipped with acoustic sensors. The underwater sensing platforms represented by UWSNs have received increased attention across a number of applications in recent years due to their advantages over wired sensor array with regard to adaptability and scalability [33]-[35]. However, the complex underwater environment exacerbates the aforementioned challenges, with channel fading being more pronounced in UWSNs due to factors such as water absorption, temperature gradients, and turbulent water motion [36], [37]. Therefore, the proposed DUKF-Fc algorithm will be employed to cope with the challenges posed by channel fading. Further, due to the fact that the constraints of energy resources on the underwater sensors will degrade the performance of the estimator, a trade-off between tracking accuracy and energy consumption using DUKF-Fc is investigated with necessity.

The remainder of this paper is organized as follows: Section II formulates the problem, incorporating the models adopted and the effects of channel fading, while the DUKF-Fc algorithm is presented in Section III. Subsequently in Section IV, the stability of DUKF-Fc is analyzed. Experimental simulations and corresponding results are provided in Section V. Finally, Section VI concludes the paper.

The notations utilized in this paper are listed as follows:

 \begin{tabular}{>{\raggedright\arraybackslash}m{1.5cm}>{\raggedright\arraybackslash}m{6.5cm}}
	$\bA^{-1}$ & The inverse matrix of $\bA$ \\
	$\bA^{\tp}$ & The transposed matrix of $\bA$ \\
	$\bA \prec \bB$ & $\bA-\bB$ is negative definite \\
	$\bA \preceq \bB$ & $\bA-\bB$ is negative semi-definite 
\end{tabular}

\begin{tabular}{>{\raggedright\arraybackslash}m{1.5cm}>{\raggedright\arraybackslash}m{6.5cm}}
	$\mathbb{E}\{\cdot\}$ & The expectation \\
	$\mathbb{P}(\cdot)$ & The probability  \\
	$\mathcal{R}(\sigma^2)$ & The Rayleigh distribution with variance ${\sigma^2}$\\ 
	$\mathcal{N}(\bsl{\mu}, \bR)$ & The Gaussian distribution with mean $\bsl{\mu}$ and covariance $\bR$ \\
	$|\mathbf{S}|$ & The cardinality of set $\mathbf{S}$ \\
	$\bI_n$ & The identity matrix with dimensions $n \times n$ \\
	$\mathbf{0}_n$ & The null vector with dimensions $n \times 1 $ \\
	$\mathbb{N}^+$ & The set of positive integers \\
	$\mathbb{R}^{m \times n}$ & The real matrix space with dimensions $m \times n$  \\
	$(\bA)(\cdot)^{\tp}$ & The simplified notation of $(\bA)(\bA)^{\tp}$  \\
	$(\bA)^{\tp}(\cdot)$ & The simplified notation of $(\bA)^{\tp}(\bA)$  \\
	$(\bA)\bB(\cdot)^{\tp}$ & The simplified notation of $(\bA)\bB(\bA)^{\tp}$ \\
	$(\bA)^{\tp}\bB(\cdot)$ & The simplified notation of $(\bA)^{\tp}\bB(\bA)$
\end{tabular}

\section{Problem Formulation}

\IEEEpubidadjcol

\subsection{Communication Graph}

The communication topology is modeled by a directed graph $\mathcal{G} = (\mathcal{V}, \mathcal{E})$, $\mathcal{V} = \{1, 2, \ldots, N\}$ being the set of nodes in WSNs, and $\mathcal{E} \subseteq \mathcal{V} \times \mathcal{V}$ representing the set of communication links among nodes. A directed link from node $i$ to node $j$ is denoted by $(i, j)$, indicating that node $j$ can transmit information to node $i$. In this context, node $j$ is considered an in-neighbor of node $i$, while node $i$ is considered an out-neighbor of node $j$. Note that a node is considered both as its own in-neighbor and out-neighbor in this paper. Therefore, the in-neighbors and out-neighbors of node $i$ can be denoted by the sets ${\iN_i} = \{ j \in \mathcal{V} \vert (j,i) \in \mathcal{E} \} \cup \{i\}$ and ${\oN_i} = \{ j \in \mathcal{V} \vert (i,j) \in \mathcal{E} \} \cup \{i\}$, respectively.

\subsection{System Dynamics}

The dynamical system for tracking maneuvering target can be described by the following equations:
\begin{align}
	&\bx_{k+1} = \bF_k\bx_k+ \bw_k, \label{TargetFunction}  \\
	&\by^i_k = \bh^i_k(\bx_k) + \bv^i_k, \ i \in \cV, \label{MeasurementFunction} 
\end{align}

\noindent where $\bx_k \in \mathbb{R}^{n \times 1}$ and $\by^i_k \in \mathbb{R}^{m \times 1}$ are the target state vector and the sensor observation vector of node $i$ at time instant $k$. $\bF_k \in \mathbb{R}^{n \times n}$ denotes the state transition matrix and $\bh^i_k: \mathbb{R}^{n \times 1} \rightarrow \mathbb{R}^{m \times 1}$ denotes the nonlinear measurement function of node $i$ valid for time $k$. The process noise and measurement noise are represented by terms $\bw_k$ and $\bv^i_k$, which are typically assumed to be mutually uncorrelated Gaussian noise with zero-mean and covariance matrices $\bQ_k, \bR^i_{v, k}$, i.e., $\bw_k \sim \cN(\mathbf{0}_n, \bQ_k), \bv^i_k \sim \cN(\mathbf{0}_m, \bR^i_{v, k})$.

\subsection{Effects of Channel Fading}
In general, the wireless channel fading can cause different types of errors in the received signal, such as amplitude distortion, phase distortion, and multi-path distortion [10], [11]. These distortions can cause errors in the demodulation and decoding process, consequentially leading to transmission failure and measurement fluctuation.

\subsubsection{Transmission Failure}
The transmission failure can occur when the receiver's signal-to-noise ratio (SNR) is too low to detect the packet's information due to channel fading. In this paper, the commonly utilized block-fading channel
model is adopted and the block length is assumed to equal the packet transmission time [38]. To model the transmission failure, the binary indicators $\{ \gam^{ij}_k \}_{i \in \cV, j \in \oN_i}$ are introduced, expressed as:
\begin{align*}
	\gam^{ij}_k = \begin{cases}
		1, &  
		\begin{aligned}
			&\mbox{the packet delivered from node $i$ to node $j$} \\
			&\mbox{successfully arrives}
		\end{aligned} \\
		0, &\mbox{otherwise} \\
	\end{cases},
\end{align*}

\noindent and the corresponding probability is given as $ \mathbb{P}( \gam^{ij}_k = 1 ) = q^{ij}_k, \ q^{ij}_k \in [0, 1]$. According to [9], [13], $q^{ij}_k$ is determined by the state of communication channel between node $i$ and node $j$ at time instant $k$, following
\begin{align}
	q^{ij}_k = \left( 1 - \beta_i( u^{ij}_k \cdot g^{ij}_k ) \right)^{l^{ij}_k},
	\label{ppp}
\end{align}

\noindent where $u^{ij}_k$ denotes the peak power level, $g^{ij}_k$ denotes the channel power gain, and $l^{ij}_k$ denotes the number of bits contained in the packet. $\beta_i: [0, \infty) \rightarrow [0, 1]$ represents the function of bit-error rate (BER), which depends on the modulation method employed by node $i$.

\subsubsection{Measurement Fluctuation}
The measurement fluctuation can occur when the returned signal from the target is weak due to channel fading and consequently, the receiver fails to correctly decode the packet and obtains biased measurements. Based on the block-fading assumption, we introduce $\vartheta^i_k$ as the fading coefficient, which is assumed to be Rayleigh distributed with variance $(\sigma^i_{\vartheta})^2$, i.e., $\vartheta^i_k \sim \cR(\sigma^i_{\vartheta})$. The fluctuated measurements are given by:
 \begin{align}
 	\bz^i_k &= \vartheta^i_k \by^i_k + \bn^i_k, \label{FadingMeasurement} 
 \end{align}

\noindent where $\bn^ i_k \in \mathbb{R}^{m \times 1}$ denotes the additional measurement noise (e.g., quantization noise), following $\cN(\mathbf{0}_m, \bR^i_{n, k})$. 

In practical applications, the fading coefficients are difficult to be precisely known and can only be estimated by utilizing channel estimation techniques at every time instant $k$. Similar to [14], the estimated fading coefficient is given by
\begin{align}
	\hat{\vartheta}^i_k = (1 + \varepsilon^i_k) \cdot \vartheta^i_k, \label{EstimatedChannelGain}
\end{align}

\noindent where $\varepsilon^i_k$ denotes the relative estimation error, which is assumed to be distributed in  $\cN(0, (\sigma^{i}_{\varepsilon})^2 )$ with truncated area $[-\Delta^i_{\varepsilon}, \Delta^i_{\varepsilon}], \ \Delta^i_{\varepsilon} \in (0, 1)$.

Substituting (5) into (4), it yields that
\begin{align}
	\bz^i_k &= \hat{\vartheta}^i_k\bh^i_k(\bx_k) + \left( \vartheta^i_k\bv^i_k + \bn^i_k - \varepsilon^i_k \vartheta^i_k \bh^i_k(\bx_k) \right).
\end{align}

Further letting $\nu^i_k = \vartheta^i_k\bv^i_k + \bn^i_k - \varepsilon^i_k \vartheta^i_k \bh^i_k(\bx_k) $, the fluctuated measurement becomes
\begin{align}
	\bz^i_k = \hat{\vartheta}^i_k\bh^i_k(\bx_k) + \nu^i_k.
	\label{FadingMeasurementFunctionModified}
\end{align}

\noindent where $\nu^i_k$ has zero mean with covariance
\begin{align}
	&{\bR}^i_{\nu,k} =  \Delta{\bR}^i_{\nu,k} + 2(\sigma^i_{\vartheta})^2\bR^i_{v,k} + \bR^i_{n,k}, \label{cc1} \\ 
	& \Delta{\bR}^i_{\nu,k} = 2(\sigma^i_{\vartheta}\sigma^i_{\varepsilon})^2 \left( 1- \dfrac{{\Delta^i_{\varepsilon}} \cdot \phi({\Delta^i_{\varepsilon}}/{\sigma^i_{\varepsilon}})}{{\sigma^i_{\varepsilon}} \cdot\Phi({\Delta^i_{\varepsilon}}/{\sigma^i_{\varepsilon}})-{\sigma^i_{\varepsilon}}/2}\right) \notag \\
	&\qquad \qquad \times  \left(\bh^i_k(\bx_{k})\bh^{i\tp}_k(\bx_{k})\right), \label{cc2}
\end{align}

\noindent under the assumption that all terms corresponding to noise are mutually uncorrelated, while
$$
\phi(\xi) = \dfrac{\exp(-\xi^2/2)}{\sqrt{2\pi}}, \ \Phi(\xi) = \int_{-\infty}^{\xi}\dfrac{\exp(-t^2/2)}{\sqrt{2\pi}} \mathrm{d}t.
$$

\section{Distributed Unscented Kalman Filter under Channel Fading}

In this section, the distributed unscented Kalman filter under channel fading named DUKF-Fc is developed to dispose the state estimation problem with impacts from wireless channel fading. Before analysis, the following assumptions are introduced.
\begin{assumption}
	For $k \in \mathbb{N}^+, i \in \cV$, $\gam^{ji}_k \equiv 1$ for $j = i$ and the indicators $\{ \gam^{ji}_k \}_{j \in \iN_i}$ is available at node $i$. (This can be achieved by utilizing techniques like error detection coding at the gateway [9], [39].)
\end{assumption}
\begin{assumption}
	For $k \in \mathbb{N}^+, i \in \cV$, the terms $\vartheta^i_k, \varepsilon^i_k, \bw^i_k, \bv^i_k, \bn^i_k$ and initial state $\bx_0$ are mutually uncorrelated. Also, $\vartheta^i_k$ is independent of $\vartheta^j_s$ when $i \neq j, k \neq s$ and the same condition holds for $\varepsilon, \bw, \bv, \bn$.
\end{assumption}

The DUKF-Fc is established on the standard UKF [40]. As for node $i$, given the estimate $\hbx^i_{k-1|k-1}$ and its covariance $\hbP^i_{k-1|k-1}$ at time instant $k$, $2n+1$ sigma points are sampled through
\begin{align}
	&\cX^{i, 0}_{k-1|k-1} =  \hbx^i_{k-1|k-1}, \notag \\
	&\cX^{i, s}_{k-1|k-1} =  \hbx^i_{k-1|k-1} + \left( \sqrt{(n+\kappa)\hbP^i_{k-1|k-1}}\right)_{(s)}, \notag \\
	&\cX^{i, s+n}_{k-1|k-1} =  \hbx^i_{k-1|k-1}- \left( \sqrt{(n+\kappa)\hbP^i_{k-1|k-1}}\right)_{(s)},
	\label{SigmaPoints}
\end{align}

\noindent where  $s = 1,\ldots, n$ and $(\cdot)_{(s)}$ represents the $s$th column of the corresponding matrix. $\kappa$ denotes the scaling factor.

Propagating $\{ \cX^{i, s}_{k-1|k-1} \}^{2n}_{s=0}$ through (1), the predicted state $\hbx^i_{k|k-1}$ and its covariance $\hbP^i_{k|k-1}$ can be calculated by combining the weighted propagated sigma points:
\begin{align}
	&\cX^{i, s}_{k|k-1} = \bF_{k-1}\cX^{i, s}_{k-1|k-1}, \ s = 0, \ldots, 2n, \label{Prediction1} \\
	&\hbx^i_{k|k-1} = \sum_{s = 0}^{2n} w_s \cX^{i, s}_{k|k-1}, \label{Prediction2}\\
	&\hbP^i_{k|k-1} = \sum_{s = 0}^{2n} w_s (\cX^{i, s}_{k|k-1}-\hbx^i_{k|k-1})(\cdot)^\tp   + \bQ_{k-1}, \label{Prediction3} 
\end{align}

\noindent where the weights are given by
\begin{align}
	w_s = \begin{cases}
		\kappa/(n+\kappa), & s = 0 \\
		1/(2n+2\kappa), & s = 1, \ldots, 2n \\
	\end{cases}.
\end{align}

Similarly, the predicted measurements $\hat{\bz}^{i}_k$ can be obtained by mapping the propagated sigma points $\{ \cX^{i, s}_{k|k-1} \}^{2n}_{s=0}$ through (7):
\begin{align}
	&\xi^{i, s}_k = \hat{\vartheta}^i_k\bh(\cX^{i, s}_{k|k-1}), \ s = 0, \ldots, 2n, \\
	&\hat{\bz}^{i}_k = \sum_{s = 0}^{2n}w_s \xi^{i, s}_k.
	\label{Prediction4} 
\end{align}

Then the measurement covariance $\hbP^i_{zz, k}$ and the state-measurement cross-covariance $\hbP^i_{xz, k}$ can be obtained by:
\begin{align}
	&\hbP^i_{zz, k} = \sum_{s = 0}^{2n} w_s (\xi^{i, s}_{k}-\hat{\bz}^{i, s}_k)(\cdot)^\tp + \hat{\bR}^i_{\nu, k}, \label{Update1} \\
	&\hbP^i_{xz, k} = \sum_{s = 0}^{2n} w_s (\cX^{i, s}_{k|k-1}-\hbx^i_{k|k-1})(\xi^{i, s}_{k}-\hat{\bz}^{i, s}_k)^\tp, \label{Update2} 
\end{align}

\noindent where the predicted covariance of $\nu^i_k$ is computed by substituting $\hbx^i_{k|k-1}$ to $\bx_k$ in (8) and (9), namely
\begin{align}
	&\hat{\bR}^i_{\nu,k} =  \Delta \hat{\bR}^i_{\nu,k} + 2(\sigma^i_{\vartheta})^2\bR^i_{v,k} + \bR^i_{n,k},  \\ 
	& \Delta \hat{\bR}^i_{\nu,k} = 2(\sigma^i_{\vartheta}\sigma^i_{\varepsilon})^2 \left( 1- \dfrac{{\Delta^i_{\varepsilon}} \cdot \phi({\Delta^i_{\varepsilon}}/{\sigma^i_{\varepsilon}})}{{\sigma^i_{\varepsilon}} \cdot\Phi({\Delta^i_{\varepsilon}}/{\sigma^i_{\varepsilon}})-{\sigma^i_{\varepsilon}}/2}\right) \notag \\
	&\qquad \qquad \times  \left(\bh^i_k(\hbx^i_{k|k-1})\bh^{i\tp}_k(\hbx^i_{k|k-1})\right), \label{Prediction5} 
\end{align}

Recalling the standard Kalman filter, the estimated state $\hbx^i_{k|k}$ and its covariance $\hbP^i_{k|k}$ are updated by
\begin{align}
	&\hbx^i_{k|k} = \hbx^i_{k|k-1} + \bK^i_{k}(\bz^i_k - \hat{\bz}^i_k),  \label{StateUpdate}\\
	&\hbP^i_{k|k} = \hbP^i_{k|k-1} -  \bK^i_{k} \hbP^i_{zz, k}\bK^{i\tp}_{k}, \label{CovarianceUpdate}
\end{align}

\noindent and the filter gain is derived by
\begin{align}
\bK^i_{k} = \hbP^i_{xz, k}(\hbP^i_{zz, k})^{-1}. \label{FilterGain}
\end{align}

Further letting 
\begin{align}
	\cH^i_{k} &= \hbP^{i\tp}_{xz, k}(\hbP^i_{k|k-1})^{-1}, \label{Update3} \\
	\cR^i_{k}  &= \hbP^i_{zz, k} -  \hbP^{i\tp}_{xz, k}(\hbP^i_{k|k-1})^{-1}\hbP^{i}_{xz, k}, \label{Update4}
\end{align}

\noindent we can formalize the estimator described by (21)-(23) to the information filter (IF):
\begin{align}
	&(\hbP^i_{k|k})^{-1} = (\hbP^i_{k|k-1})^{-1} + \cH^{i\tp}_{k} ( \cR^i_{k} )^{-1}\cH^{i}_{k}, \\
	&\hbx^i_{k|k} = \hbx^i_{k|k-1} + \hbP^i_{k|k}  \cH^{i\tp}_{k} ( \cR^i_{k} )^{-1}(\bz^i_k - \hat{\bz}^i_k),
\end{align}

\noindent and accordingly, it follows from [28] that the multi-sensor case is formulated as
\begin{align}
&(\hbP^i_{k|k, l})^{-1} = (\hbP^i_{k|k-1})^{-1} + \sum_{j \in \iN_i} \gamma^{ji}_k \cH^{j\tp}_{k} ( \cR^j_{k} )^{-1}\cH^{j}_{k}, \label{Update5}\\
&\hbx^i_{k|k, l} = \hbx^i_{k|k-1} + \hbP^i_{k|k, l} \sum_{j \in \iN_i}\gamma^{ji}_k \cH^{j\tp}_{k} ( \cR^j_{k} )^{-1}(\bz^j_k - \hat{\bz}^j_k), \label{Update6}
\end{align}

\noindent where $\hbx^i_{k|k, l}$ is the local estimate at node $i$ and $\hbP^i_{k|k, l}$ its covariance. Subsequently, the global estimate and its covariance at node $i$, indicated by $\hbx^i_{k|k}, \hbP^i_{k|k}$, are obtained after the diffusion strategy, given by
\begin{align}
	&\hbx^i_{k|k} = \sum_{j \in \iN_i} \bC_k^{(i, j)} \hbx^j_{k|k, l}, \label{Update7}\\
	&\hbP^i_{k|k} = \sum_{j \in \iN_i} \bC_k^{(i, j)} \hbP^j_{k|k, l}, \label{Update8}
\end{align}

\noindent where $\bC_k$ denotes the diffusion matrix at time instant $k$ and $\bC_k^{(i, j)}$ its $(i, j)$ element. According to [41], $\bC_k$ possesses the following properties:
$$
\mathbf{1}_n \cdot \bC_k = \mathbf{1}_n, \ 
\bC_k^{(i, j)} = 0 \ \mbox{if} \ j \notin \iN_i, \ \bC_k^{(i, j)} \geq 0 \ \forall i, j \in \cV.
$$

Considering the possibility of transmission failure when exchanging local estimates, we introduce the indicators $\{ \gam^{ji}_{k+} \}_{j \in \iN_i}$, which maintain the same Assumptions 1-2 as $\{ \gam^{ji}_k \}_{j \in \iN_i}$. In this paper, the Metropolis's rule [42] is adopted and the diffusion matrix is defined as:
\begin{align}
	\bC_k^{(i, j)} &= \begin{cases}
		\dfrac{\gam^{ji}_{k+}}{\max \{ |\iN_i| , |\iN_j| \}}, & 
        j \in \iN_i \ \mbox{and} \ j \neq i  \\
		1 - \sum_{j \in \cV \backslash \{i\}} \bC_k^{(i, j)}, &j = i  \\
		0, &j \notin \iN_i.
	\end{cases}
\label{DiffusionMatrix}
\end{align}

For each node $i \in \cV$, the procedure described above is conducted in parallel, generating a list of estimates $\{ \hbx^i_{k|k} \}_{i \in \cV}$ at every time instant $k$. The proposed DUKF-Fc is summarized as algorithm 1 given below.

\begin{algorithm}[H]
	\caption{DUKF-Fc}\label{DUKF}
	\begin{algorithmic}
		\STATE 
		\STATE \hspace{0.5cm} Consider the system dynamics described by (1)-(2) and (7). For $\forall i \in \cV$, start with $\hbx^i_{0|0} = \bx_0$ and $\hbP^i_{0|0} = \bP_0$ and for $k \geq 1$, iterate:
		\STATE 
		\STATE {\textbf{Step 1}: Sampling and Prediction}
		\STATE \hspace{0.5cm} Sample the sigma points through (10) and calculate the predictions $\hbx^i_{k|k-1}, \hbP^i_{k|k-1}, \hat{\bz}^{i}_k$ and  $\hat{\bR}^i_{\nu,k}$ by (12), (13), (16) and (20), respectively.
		\STATE
		\STATE {\textbf{Step 2}: Local Estimation}
		\STATE \hspace{0.5cm} Calculate the covariance matrices $\hbP^i_{zz, k}, \hbP^i_{xz, k}$ through (17)-(18) and accordingly, compute $\cH^{i}_{k}, \cR^i_{k}$ by (24)-(25).
		 \STATE \hspace{0.5cm} Encapsulate the information pair $( \cH^{i\tp}_{k} ( \cR^i_{k} )^{-1}\cH^{i}_{k}, \cH^{i\tp}_{k} ( \cR^i_{k} )^{-1}(\bz^i_k - \hat{\bz}^i_k) )$ and communicate it with out-neighbors to update the local estimates $\hbx^i_{k|k, l}, \hbP^i_{k|k, l}$ using (28) and (29).
		\STATE
		\STATE {\textbf{Step 3}: Global Estimation}
		\STATE \hspace{0.5cm} Exchange $\hbx^i_{k|k, l}, \hbP^i_{k|k, l}$ with out-neighbors and calculate the global estimates $\hbx^i_{k|k}, \hbP^i_{k|k}$ by (30)-(32).
	\end{algorithmic}
	\label{alg1}
\end{algorithm}

\section{Performance Analysis of the Proposed Algorithm}

The performance of the proposed DUKF-Fc is analyzed in this section, including the boundedness of the error covariance and the stochastic stability of the filter. For ease of analysis, we adopt the approach presented in [43], which employs the first-order Taylor series expansion to simplify the system dynamics. The system dynamics described by (1) and (7) thus becomes
\begin{align}
	&\bx_{k+1} = \bF_k\bx_k+ \bw_k, \\
	&\bz^i_k = \hat{\vartheta}^i_k \bsl{\beta}^i_k \bH^i_k \bx_k + \nu^i_k.
\end{align}

\noindent where $\bsl{\beta}^i_k = \diag\{ {\beta}^{i, 1}_k, \ldots, {\beta}^{i, m}_k  \}$ denotes an unknown diagonal matrix for compensating the linearization errors and
 $$\bH^i_{k} =  \left.\dfrac{\partial \bh^i_k( \bx_k ) }{\partial \bx_k} \right|_{\hbx^i_{k|k-1}}.$$

Then the matrices $\hbP^i_{k|k-1}, \hbP^i_{zz, k}$ and $\hbP^i_{xz, k}$ can be rearranged as:
\begin{align}
	&\hbP^i_{k|k-1} = \bF_{k-1} \hbP^i_{k-1|k-1} \bF_{k-1}^{\tp} + \bQ_{k-1}, \label{PredictionCovariance2}\\
	&\hbP^i_{zz, k} = (\hat{\vartheta}^i_k)^2 \bsl{\beta}^i_k \bH^i_k \hbP^i_{k|k-1} \bH^{i\tp}_k\bsl{\beta}^i_k + \hat{\bR}^i_{\nu, k}, \\
	&\hbP^i_{xz, k} = \hat{\vartheta}^i_k \hbP^i_{k|k-1}  \bH^{i\tp}_k\bsl{\beta}^i_k,
\end{align}

\noindent and accordingly,
\begin{align}
	&\cH^i_{k} = \hbP^{i\tp}_{xz, k}(\hbP^i_{k|k-1})^{-1} =  \hat{\vartheta}^i_k \bsl{\beta}^i_k\bH^{i}_k, \label{H}\\
	&\cR^i_{k}  = \hbP^i_{zz, k} -  \cH^i_k\hbP^{i}_{xz, k} = \hat{\bR}^i_{\nu, k}. \label{R}
\end{align}

Substituting (38)-(39) to (28)-(29), the update law becomes
\begin{align}
	&(\hbP^i_{k|k, l})^{-1} = (\hbP^i_{k|k-1})^{-1} + \bS^i_k \label{Pupdate}\\
	&\hbx^i_{k|k, l} = \hbx^i_{k|k-1} + \hbP^i_{k|k, l}( \bq^i_k - \bS^i_k \hbx^i_{k|k-1} )
\end{align}

\noindent where
\begin{align}
	&\bS^i_k =  \sum_{j \in \iN_i}(\hat{\vartheta}^j_k)^2 \gamma^{ji}_k\bH^{j\tp}_k\bsl{\beta}^j_k  ( \hat{
	\bR}^j_{\nu, k} )^{-1}\bsl{\beta}^j_k\bH^{j}_{k}, \label{sss}\\
	&\bq^i_k = \sum_{j \in \iN_i}\hat{\vartheta}^j_k \gamma^{ji}_k \bH^{j\tp}_k\bsl{\beta}^j_k  ( \hat{\bR}^j_{\nu, k} )^{-1}\bz^j_k.
\end{align}

Before proceeding, the following assumptions and lemmas are introduced with necessity.

\begin{assumption}
	There exist real numbers $ \ul{f}, \ul{h}, \ul{\beta},\ul{\vartheta},\ul{q}, \ul{r}_v, \ul{r}_n$ and $\ol{\beta}, \ol{f} , \ol{h}, \ol{\beta},\ol{\vartheta},\ol{q}, \ol{r}_v, \ol{r}_n$ such that for $i \in \cV, k \in \mathbb{N}^+$,
	\begin{align*}
		&\begin{cases}
			\ul{f}^2\bI_n \preceq \bF_k\bF^{\top}_k \preceq \ol{f}^2\bI_n \\
			\ul{h}^2\bI_m \preceq \bH^i_k\bH^{i\top}_k \preceq \ol{h}^2\bI_m \\
			\ul{\beta}^2\bI_m \preceq \bsl{\beta}^i_k\bsl{\beta}^{i\top}_k \preceq \ol{\beta}^2\bI_m
		\end{cases} \mbox{\&} \quad \begin{cases}
		\ul{\vartheta} \leq \vartheta^i_k \leq \ol{\vartheta} \\
		\ul{q}\bI_n \preceq \bQ_k \preceq \ol{q}\bI_n \\
		\ul{r}_v \bI_m \preceq {\bR}^i_{v, k} \preceq \ol{r}_v\bI_m \\
		\ul{r}_n \bI_m \preceq {\bR}^i_{n, k} \preceq \ol{r}_n\bI_m 
	\end{cases}.
	\end{align*}

Also, it is assumed that the covariance corresponding to channel estimation error is bounded for $\forall i \in \cV, k \in \mathbb{N}^+$, i.e., 
$$
\ul{r}_{\nu} \bI_m \preceq  \Delta \hat{\bR}^i_{\nu,k} \preceq \ol{r}_{\nu} \bI_m,
$$

\noindent yielding 
$$
\ul{r} \bI_m \preceq  \hat{\bR}^i_{\nu,k} \preceq \ol{r} \bI_m.
$$

\noindent where $\ul{r} = 2( \ul{\sigma}^i_{\vartheta} )^2 \ul{r}_v +  \ul{r}_n + \ul{r}_{\nu}, \ol{r} = 2( \ol{\sigma}^i_{\vartheta} )^2 \ol{r}_v +  \ol{r}_n + \ol{r}_{\nu}$ with $\ul{\sigma}^i_{\vartheta} = \min_{i \in \cV} \sigma^i_{\vartheta}, \ol{\sigma}^i_{\vartheta} = \max_{i \in \cV} \sigma^i_{\vartheta}$.

\end{assumption}

\begin{assumption}
	The weight matrix $\bC_k$ is row-stochastic and primitive for $k \in \mathbb{N}^+$.
\end{assumption}

\begin{lemma}[44]
	Given real numbers $0 \leq \eta_1 < 1, \eta_2 > 0,\ul{v}$ and  $\ol{v}$, the stochastic process $\bV_k(\bx_k)$ is exponentially bounded in mean square, i.e.,
	\begin{align*}
		\mathbb{E}\{ \Vert \bx_k  \Vert^2 \} \leq \dfrac{\ol{v}}{\ul{v}} \mathbb{E}\{ \Vert \bx_0  \Vert^2 \} \eta_1^k +  \dfrac{\eta_2}{\ul{v}} \sum_{i=1}^{k-1}\eta_1^i
	\end{align*}
	
	\noindent when
	\begin{align}
		& \mathbb{E}\{ \bV_{k+1}(\bx_{k+1}) | \bx_{k} \} \leq \eta_1 \bV_{k}(\bx_k) + \eta_2, \label{c1}\\
		& \ul{v} \Vert \bx_k \Vert^2 \leq \bV_k(\bx_k) \leq \ol{v} \Vert \bx_k \Vert^2. \label{c2}
	\end{align}
\end{lemma}

\begin{lemma}[45]
	The following inequality holds for any two symmetric positive definite matrices $\bA$ and $\bB$:
	\begin{align*}
		 \bA^{-1} - \bA^{-1} \bB \bA^{-1} \preceq  (\bA + \bB)^{-1}.
	\end{align*}
\end{lemma}

\begin{lemma}[46]
	Given integer $N \in \mathbb{N}^+$, a set of positive definite matrices $\{\bM_i\}^N_{i=1}$ and a set of vectors $\{\bv_i\}^N_{i=1}$. The following inequality holds
	\begin{align*}
		\left( \sum_{i = 1}^{N} 
		\bM_i \bv_i \right)^{\tp} \left( \sum_{i = 1}^{N} 
		\bM_i \right)^{-1} \left( \sum_{i = 1}^{N} 
		\bM_i \bv_i \right) \preceq \sum_{i = 1}^{N} \bv^{\tp}_i \bM_i \bv_i 
	\end{align*}
\end{lemma}

\begin{lemma}[28]
	Assuming that $\{\bA_i\}, \{\bB_i\}$ are two invertible positive definite matrices sequences satisfying
	\begin{align*}
		&\ul{a}\bI \preceq \bA_i \preceq \ol{a}\bI, \\
		&\ul{b}\bI \preceq \bB_i \preceq \ol{b}\bI
	\end{align*}
	
	\noindent for $i \in \mathbb{N}^+$. There exist two positive scalars $0 < \theta_1 < \theta_2 < 1$ such that
	$$
	\theta_1 \bA_i^{-1} \preceq (\bA_i + \bB_i)^{-1} \preceq \theta_2 \bA_i^{-1}
	$$
	
	\noindent where
	$$
	\theta_1 = \ul{a} / (\ul{a}+\ol{b}) , \ \theta_2 = \ol{a} / (\ol{a}+\ul{b}) .
	$$
\end{lemma}

\begin{lemma}[47]
	Assuming that $\{a_i\}, \{b_i\}$  are two positive numbers sequences with length $N$, then
	$$
	{\left(\sum_{i=1}^{N}a_i\right)}/{\left(\sum_{i=1}^{N}b_i\right)} \leq \max_{i = 1, \ldots, N} \dfrac{a_i}{b_i},
	$$
	
	\noindent and the equality holds if and only if all the ratios $a_i/b_i$ are equal.
\end{lemma}

\begin{theorem}
	Under Assumptions 1-3, there exists lower and upper bounds $\ul{p}$ and $\ol{p}$ such that the global estimation error covariance $\hbP^i_{k|k}$ of the modified DUKF described in algorithm 1 satisfies 
	\begin{align}
		\ul{p} \bI_n \preceq \hbP^i_{k|k} \preceq \ol{p} \bI_n, \ \mbox{for} \ i \in \cV,
	\end{align}

\noindent where
\begin{align}
	& \ul{p} = \left( \ul{q}^{-1} + \dfrac{(1 + \ol{\Delta}_{\varepsilon})^2 N \ol{h}^2  \ol{\beta}^2 \ol{\vartheta}^2}{\ul{r} } \right)^{-1}, \label{lb} \\
	& \ol{p} = \dfrac{\ol{h}^2 \ol{r} }{(1-\ol{\Delta}_{\varepsilon})^2 \ul{h}^4  \ul{\beta}^2 \ul{\vartheta}^2 },\label{ub}
\end{align}
\end{theorem}

\noindent with $\ol{\Delta}_{\varepsilon} = \max_{i \in \cV} \Delta^i_{\varepsilon}$.

\textit{Proof:} It can be obtained from (42) that
\begin{align}
	\bS^i_k &=  \sum_{j \in \iN_i}(\hat{\vartheta}^j_k)^2 \gamma^{ji}_k \bH^{j\tp}_k\bsl{\beta}^j_k  ( \hat{\bR}^j_{\nu, k} )^{-1}\bsl{\beta}^j_k\bH^{j}_{k} \notag \\
	&\preceq  \sum_{j \in \iN_i}(\hat{\vartheta}^j_k)^2  \bH^{j\tp}_k\bsl{\beta}^j_k  ( \hat{\bR}^j_{\nu, k} )^{-1}\bsl{\beta}^j_k\bH^{j}_{k} \notag \\
	&\preceq \dfrac{(1 + \ol{\Delta}_{\varepsilon} )^2\ol{h}^2 \ol{\beta}^2}{\ul{r} }  \sum_{j \in \iN_i} (\vartheta^j_k)^2 \label{lower1}
\end{align}

\noindent where $\ol{\Delta}_{\varepsilon} = \max_{j \in \cV} \Delta^j_{\varepsilon}$. Also, one can derive from (31) and (40) that
\begin{align}
	\hbP^i_{k|k} &=  \sum_{j \in \iN_i} \bC_k^{(i, j)}  \left((\bF_{k-1} \hbP^j_{k-1|k-1} \bF_{k-1}^{\tp} + \bQ_{k-1})^{-1} + \bS^j_k \right)^{-1} \notag \\
	&\succeq  \sum_{j \in \iN_i} \bC_k^{(i, j)}  \left((\bQ_{k-1})^{-1} + \bS^i_k\right)^{-1} \label{lower2}
\end{align}

Substituting (49) to (50) leads to
\begin{align}
	\hbP^i_{k|k} 
	&\succeq   \sum_{j \in \iN_i} \bC_k^{(i, j)} \Bigg( \ul{q}^{-1} + \dfrac{(1 + \ol{\Delta}_{\varepsilon})^2\ol{h}^2 \ol{\beta}^2}{\ul{r} } \sum_{l \in \iN_j} (\vartheta^l_k)^2 \Bigg)^{-1} \bI_n \notag \\
	&\succeq \left( \ul{q}^{-1} + \dfrac{(1 + \ol{\Delta}_{\varepsilon})^2 \ol{h}^2  \ol{\beta}^2}{\ul{r} } \sum_{i \in \cV} (\vartheta^i_k)^2 \right)^{-1} \bI_n \notag \\
	&\succeq \left( \ul{q}^{-1} + \dfrac{(1 + \ol{\Delta}_{\varepsilon})^2 N \ol{h}^2  \ol{\beta}^2 \ol{\vartheta}^2}{\ul{r} } \right)^{-1} \bI_n \notag \\
	&:= \ul{p} \bI_n \ ,
\end{align}

\noindent which gives the lower bound in (47).

By applying the Woodbury identity of matrix inverse, it follows from  (40) and (42)  that
\begin{align}
	\hbP^i_{k|k, l} &=  \hbP^i_{k|k-1}  - \Bigg(\hbP^i_{k|k-1} \bigg( \hbP^i_{k|k-1}  \notag \\
	&\quad + \Big(\sum_{j \in \iN_i}(\hat{\vartheta}^j_k)^2 \gamma^{ji}_k (\bsl{\beta}^j_k \bH^{j}_k)^{\tp} ( \hat{\bR}^j_{\nu, k} )^{-1}(\cdot)\Big)^{-1} \bigg)\hbP^i_{k|k-1} \Bigg) \notag  \\
	&\preceq \hbP^i_{k|k-1}  - \Bigg(\hbP^i_{k|k-1} \bigg( \hbP^i_{k|k-1}  \notag \\
	&\quad + \Big( (\hat{\vartheta}^i_k)^2 (\bsl{\beta}^i_k \bH^{i}_k)^{\tp} ( \hat{\bR}^i_{\nu, k} )^{-1}(\cdot)\Big)^{-1} \bigg)\hbP^i_{k|k-1} \Bigg) \notag  \\
	&= \hbP^i_{k|k-1} -  (\hat{\vartheta}^i_k)^2   (\bsl{\beta}^i_k \bH^{i}_k \hbP^i_{k|k-1})^{\tp} \notag \\
	&\qquad \times \Bigg( (\hat{\vartheta}^i_k)^2 (\bsl{\beta}^i_k \bH^i_k) \hbP^i_{k|k-1} (\cdot)^{\tp} + \hat{\bR}^i_{\nu, k} \Bigg)^{-1} (\cdot) \label{Pupper1}
\end{align}

According to Lemma 2, the second term above holds the following inequality:
\begin{align}
	&(\bsl{\beta}^i_k \bH^{i}_k \hbP^i_{k|k-1})^{\tp} \Bigg( (\hat{\vartheta}^i_k)^2 (\bsl{\beta}^i_k \bH^i_k) \hbP^i_{k|k-1} (\cdot)^{\tp} + \hat{\bR}^i_{\nu, k} \Bigg)^{-1} (\cdot)  \notag \\
	\succeq &(\hat{\vartheta}^i_k)^{-2} \hbP^i_{k|k-1} - (\hat{\vartheta}^i_k)^{-4} \Big(\bH^{i\tp}_k (\bH^i_k \bH^{i\tp}_k)^{-1}(\bsl{\beta}^i_k)^{-1}\Big) \hat{\bR}^i_{\nu, k}(\cdot). \label{Pupper2}
\end{align}

Combining (52) and (53), one can obtain the bound valid for local error covariance, given by
\begin{align}
	\hbP^i_{k|k, l} 
	&\preceq (\hat{\vartheta}^i_k)^{-2} \Big(\bH^{i\tp}_k (\bH^i_k \bH^{i\tp}_k)^{-1}(\bsl{\beta}^i_k)^{-1}\Big) \hat{\bR}^i_{\nu, k}(\cdot) \notag \\
	&\preceq \left(\dfrac{({\vartheta}^i_k)^{-2} \ol{h}^2 \ol{r} }{(1-\ol{\Delta}_{\varepsilon})^2 \ul{h}^4 \ul{\beta}^2 }
	\right) \bI_n, \label{Pupper3}
\end{align}

Going through the diffusion strategy yields
\begin{align}
	\hbP^i_{k|k} &= \sum_{j \in \iN_i} \bC_k^{(i, j)} \hbP^j_{k|k, l} \notag \\
	&\preceq  \left(\dfrac{\ol{h}^2  \ol{r} }{(1-\ol{\Delta}_{\varepsilon})^2  \ul{h}^4  \ul{\beta}^2 }
	\sum_{j \in \iN_i}  \bC_k^{(i, j)} ({\vartheta}^j_k)^{-2}\right) \bI_n  \notag \\
	&\preceq  \left(\dfrac{\ol{h}^2 \ol{r} }{(1-\ol{\Delta}_{\varepsilon})^2 \ul{h}^4  \ul{\beta}^2 \ul{\vartheta}^2 }\right) \bI_n   \notag \\
	&:= \ol{p} \bI_n , 
\end{align}

\noindent which gives the upper bound in (48).

\begin{theorem}
	 Under Assumptions 1-4, the global estimation error of the modified DUKF described in algorithm 1 is exponentially bounded in mean square for $i \in \cV$ when there exists a real number $0 < \ol{\xi} < 1 + {\ul{q}}/{\ol{f}^2\ol{p}}$, such that
	\begin{align}
		\max_{j = 1, \ldots, N} \dfrac{c^j_{k+1}}{c^j_{k}} \leq \ol{\xi} \ \ \forall k \in \mathbb{N}^+,
	\end{align}

	\noindent where $c_k^i$ represents the $i$th element of the Perron–Frobenius left eigenvector of $\bC_k$.
\end{theorem}

\textit{Proof:} Denote $
\tbx_{k|k} = [\tbx^{1\tp}_{k|k}, \ldots, \tbx^{N\tp}_{k|k}]^{\tp}
$ and $
\hbx_{k|k} = [\hbx^{1\tp}_{k|k}, \ldots, \hbx^{N\tp}_{k|k}]^{\tp}
$. According to Assumption 4, we have
\begin{align}
	0 < \ul{c} \leq c_k^i  \leq \ol{c}, \
\sum_{i \in \cV} c_k^i \bC_k^{(i, j)} = c_k^j.
\end{align}

Define the estimation error as $\tbx_{k|k} = \bx_k - \hbx_{k|k}$ and consider the following stochastic process
\begin{align}
\bV_{k}(\tbx_{k|k}) = \sum_{i \in \cV} c_k^i \tbx^{i\tp}_{k|k} (\hbP^i_{k|k})^{-1} \tbx^i_{k|k}.
\end{align}

Recalling that $ \ul{p}\bI_n \preceq \hbP^i_{k|k} \preceq \ol{p}\bI_n $, the following inequalities can be obtained immediately
\begin{align}
	\dfrac{N \ul{c}}{\ol{p}} \Vert  \tbx_{k|k} \Vert^2 \leq \bV_{k}(\tbx_{k|k}) \leq \dfrac{N \ol{c}}{\ul{p}} \Vert  \tbx_{k|k} \Vert^2,
\end{align}

\noindent which satisfies the condition (44).

Next, we turn to figure out the recursion of $\tbx^i_{k|k}$.  The error of local estimate follows
\begin{align}
	\tbx^i_{k|k, l} &= \bx_k - \hbx^i_{k|k, l} \notag \\
	&= \bx_k - \hbx^i_{k|k-1} + \hbP^i_{k|k, l}( - \bq^i_k + \bS^i_k \hbx^i_{k|k-1} )  \notag  \\
	&= ( \bI - \hbP^i_{k|k, l} \bS^i_k) \bF_{k-1} \tbx^i_{k-1|k-1} + ( \bI - \hbP^i_{k|k, l} \bS^i_k) \bw_{k-1} \notag  \\
	&\quad - \hbP^i_{k|k, l} \sum_{j \in \iN_i}\hat{\vartheta}^j_k \gamma^{ji}_k \bH^{j\tp}_k\bsl{\beta}^j_k  ( \hat{\bR}^j_{\nu, k} )^{-1}\nu^j_{k} 
\end{align}

Further combining with (30), the relationship between $\tbx^i_{k|k}$ and $\tbx^i_{k-1|k-1}$ can be derived by
\begin{align}
	\tbx^i_{k|k} &= \sum_{j \in \iN_i} \bC_k^{(i, j)} \tbx^j_{k|k, l} \notag \\
	&= \sum_{j \in \iN_i} \bC_k^{(i, j)} ( \bI - \hbP^j_{k|k, l} \bS^j_k) \bF_{k-1} \tbx^j_{k-1|k-1}  \notag \\
	&\quad+ \sum_{j \in \iN_i} \bC_k^{(i, j)} ( \bI - \hbP^j_{k|k, l} \bS^j_k) \bw_{k-1} \notag \\
	&\quad - \sum_{j \in \iN_i} \bC_k^{(i, j)}  \hbP^j_{k|k, l} \sum_{l \in \iN_j}\hat{\vartheta}^l_k \gamma^{lj}_k \bH^{l\tp}_k\bsl{\beta}^l_k  ( \hat{\bR}^l_{\nu, k} )^{-1}\nu^l_{k} \label{stability1}
\end{align}

Denote $\bH^{i\tp}_k\bsl{\beta}^i_k  ( \hat{\bR}^i_{\nu, k} )^{-1}$ as $\bD^i_k$. Substituting (61) to $\bV_{k}(\tbx_{k|k})$, we have
\begin{align}
	&\mathbb{E}\{ \bV_{k}(\tbx_{k|k}) | \tbx_{k-1|k-1} \} \notag \\ &= \mathbb{E}\{ \sum_{i \in \cV} c_k^i \tbx^{i\tp}_{k|k} (\hbP^i_{k|k})^{-1} \tbx^i_{k|k}| \tbx_{k-1|k-1} \}  \notag  \\
	&= \mathbb{E}\Bigg\{ \sum_{i \in \cV} c_k^i 
	\Big(\sum_{j \in \iN_i} \bC_k^{(i, j)} ( \bI - \hbP^j_{k|k, l} \bS^j_k) \bF_{k-1} \tbx^j_{k-1|k-1}\Big)^{\tp} \notag \\
	&\qquad \times (\hbP^i_{k|k})^{-1}(\cdot) | \tbx_{k-1|k-1}
	\Bigg\} \notag \\
	&\quad + \mathbb{E}\Bigg\{ \sum_{i \in \cV} c_k^i 
	\Big(\sum_{j \in \iN_i} \bC_k^{(i, j)} ( \bI - \hbP^j_{k|k, l} \bS^j_k) \bw_{k-1}\Big)^{\tp} \notag \\
	&\qquad \quad \times (\hbP^i_{k|k})^{-1}(\cdot) | \tbx_{k-1|k-1}
	\Bigg\} \notag  \\ 
	&\quad + \mathbb{E}\Bigg\{ \sum_{i \in \cV} c_k^i 
	\Big(\sum_{j \in \iN_i} \bC_k^{(i, j)} \hbP^j_{k|k, l} \sum_{l \in \iN_j}\hat{\vartheta}^l_k \gamma^{lj}_k \bD^l_k\nu^l_{k}\Big)^{\tp} \notag \\
	&\qquad \quad \times (\hbP^i_{k|k})^{-1}(\cdot) | \tbx_{k-1|k-1}
	\Bigg\}
\end{align}

Let $\mathbb{A}_k, \mathbb{B}_k, \mathbb{C}_k$ denote the first, the second and the third term above. Next, we will discuss each of these three terms. To begin with, we focus on the first term $\mathbb{A}_k$ and conduct the following inequalities by applying Lemma 3
\begin{align}
	\mathbb{A}_k &= \mathbb{E}\Bigg\{ \sum_{i \in \cV} c_k^i 
	\Big(\sum_{j \in \iN_i} \bC_k^{(i, j)} ( \bI - \hbP^j_{k|k, l} \bS^j_k) \bF_{k-1} \tbx^j_{k-1|k-1}\Big)^{\tp} \notag \\
	&\qquad \times (\hbP^i_{k|k})^{-1}(\cdot) | \tbx_{k-1|k-1}
	\Bigg\}  \notag \\
	&= \mathbb{E}\Bigg\{ \sum_{i \in \cV} c_k^i \Big(\sum_{j \in \iN_i} \bC_k^{(i, j)} \hbP^j_{k|k, l} (\hbP^j_{k|k-1})^{-1} \bF_{k-1} \tbx^j_{k-1|k-1}\Big)^{\tp}\notag \\
	&\qquad \times \Big( \sum_{j \in \iN_i} \bC_k^{(i, j)} \hbP^j_{k|k, l}\Big)^{-1} (\cdot) | \tbx_{k-1|k-1}
	\Bigg\} \notag  \\
	&\leq  \mathbb{E}\Bigg\{ \sum_{i \in \cV} c_k^i 
	\Big(\sum_{j \in \iN_i} \bC_k^{(i, j)} 
	\left( (\hbP^j_{k|k-1})^{-1} \bF_{k-1} \tbx^j_{k-1|k-1}\right)^{\tp} \notag \\
	&\qquad \times (\hbP^j_{k|k, l}) (\cdot)
	\Big) | \tbx_{k-1|k-1}
	\Bigg\} \notag \\
	&= \mathbb{E}\Bigg\{ \sum_{i \in \cV} c_k^i 
	\Big(\sum_{j \in \iN_i} \bC_k^{(i, j)} 
	\tbx^{j\tp}_{k-1|k-1}\bF^{\tp}_{k-1}(\hbP^j_{k|k-1})^{-1} \notag \\
	&\qquad \times (\hbP^j_{k|k, l})(\hbP^j_{k|k-1})^{-1}\bF_{k-1} \tbx^j_{k-1|k-1}
	\Big) | \tbx_{k-1|k-1}
	\Bigg\}  \label{AAA}
\end{align}

It follows form $(\hbP^i_{k|k, l})^{-1} = (\hbP^i_{k|k-1})^{-1} + \bS^i_k$ that
\begin{align}
\hbP^i_{k|k, l} \preceq \hbP^i_{k|k-1}. \label{stability2}
\end{align}

Also, applying Lemma 4 to (35), we have the following inequality
\begin{align}
	(\hbP^j_{k|k-1})^{-1} \preceq \eta \cdot \bF^{-\tp}_{k-1} (\hbP^j_{k-1|k-1})^{-1} \bF^{-1}_{k-1}.\label{stability3}
\end{align}

\noindent where $\eta = \ol{f}^2\ol{p}/ (\ul{q} + \ol{f}^2\ol{p})$. 

Substituting (64) and (65) to (63) yields
\begin{align}
	\mathbb{A}_k 
	&\leq \mathbb{E}\Bigg\{ \sum_{i \in \cV} c_k^i 
	\Big(\sum_{j \in \iN_i} \bC_k^{(i, j)} 
	(\tbx^{j\tp}_{k-1|k-1}\bF^{\tp}_{k-1}) \notag \\
	&\qquad \times  (\hbP^j_{k|k-1})^{-1} (\bF_{k-1} \tbx^j_{k-1|k-1})
	\Big) | \tbx_{k-1|k-1}
	\Bigg\} \notag \\
	&\leq \eta \mathbb{E}\Bigg\{ \sum_{j \in \cV} c_k^j
	\tbx^{j\tp}_{k-1|k-1} (\hbP^j_{k-1|k-1})^{-1} (\cdot)
	| \tbx_{k-1|k-1}
	\Bigg\} \label{k1}\\
	&\leq  \max_{j = 1, \ldots, N} \dfrac{\eta c^j_{k+1}}{c^j_{k}} \notag \\
	&\qquad \times \mathbb{E}\Bigg\{ \sum_{j \in \cV} c_{k-1}^j
	\tbx^{j\tp}_{k-1|k-1} (\hbP^j_{k-1|k-1})^{-1} (\cdot)
	| \tbx_{k-1|k-1}
	\Bigg\}  \label{kkk2} \\
	& \leq ({\eta} \cdot \ol{\xi})  \bV_{k}(\tbx_{k-1|k-1}) 
	 := \ol{\eta} \bV_{k}(\tbx_{k-1|k-1}) 
\end{align}

\noindent where the inequality (67) is obtained by applying Lemma 5 to (66). Since $0 < \ol{\xi}  < 1 + {\ul{q}}/{\ol{f}^2\ol{p}}$, it is easy to verify that $0< \ol{\eta} <1$

Subsequently, we proceed with the boundedness of terms $\mathbb{B}_k$ and $\mathbb{C}_k$. Using Lemma 3 and (64), the following inequalities can be conducted
\begin{align}
	\mathbb{B}_k &= \mathbb{E}\Bigg\{ \sum_{i \in \cV} c_k^i 
	\Big(\sum_{j \in \iN_i} \bC_k^{(i, j)} ( \bI - \hbP^j_{k|k, l} \bS^j_k) \bw_{k-1} \Big)^{\tp} \notag \\
	&\qquad \times (\hbP^i_{k|k})^{-1}(\cdot) | \tbx_{k-1|k-1}
	\Bigg\}  \notag \\
	&= \mathbb{E}\Bigg\{ \sum_{i \in \cV} c_k^i
	\Big(\sum_{j \in \iN_i} \bC_k^{(i, j)} \hbP^j_{k|k, l} (\hbP^j_{k|k-1})^{-1} \bw_{k-1} \Big)^{\tp}\notag \\
	&\qquad \times \Big( \sum_{j \in \iN_i} \bC_k^{(i, j)} \hbP^j_{k|k, l}\Big)^{-1} (\cdot) | \tbx_{k-1|k-1}
	\Bigg\} \notag  \\
	&\leq  \mathbb{E}\Bigg\{ \sum_{i \in \cV} c_k^i
	\Big(\sum_{j \in \iN_i} \bC_k^{(i, j)} 
	\bw^{\tp}_{k-1} (\hbP^j_{k|k-1})^{-1}  \notag \\
	&\qquad \times (\hbP^j_{k|k, l}) (\hbP^j_{k|k-1})^{-1} \bw_{k-1}
	\Big) | \tbx_{k-1|k-1}
	\Bigg\} \notag \\
	&\leq  \mathbb{E}\Bigg\{ \sum_{i \in \cV} c_k^i
	\Big(\sum_{j \in \iN_i} \bC_k^{(i, j)} 
	\bw^{\tp}_{k-1} (\hbP^j_{k|k, l})^{-1} (\cdot)
	\Big) | \tbx_{k-1|k-1}
	\Bigg\} \notag  \\
	&\leq  \dfrac{N\bar{c}\bar{q}}{ \ul{p}}
\end{align}

Similar manipulations holds for $\mathbb{C}_k$, which leads to
\begin{align}
	\mathbb{C}_k &= \mathbb{E}\Bigg\{ \sum_{i \in \cV} c_k^i 
	\Big(\sum_{j \in \iN_i} \bC_k^{(i, j)}  \hbP^j_{k|k, l} \sum_{l \in \iN_j}\hat{\vartheta}^l_k \gamma^{lj}_k \bD^l_k\nu^l_{k}\Big)^{\tp}\notag \\
	&\qquad \times (\hbP^i_{k|k})^{-1}(\cdot) | \tbx_{k-1|k-1}
	\Bigg\} \notag \\
	&\leq  \mathbb{E}\Bigg\{ \sum_{i \in \cV} c_k^i 
	\sum_{j \in \iN_i} \bC_k^{(i, j)} 
	(\sum_{l \in \iN_j}\hat{\vartheta}^l_k \gamma^{lj}_k \bD^l_k\nu^l_{k})^{\tp} \notag \\
	&\qquad \times (\hbP^j_{k|k, l}) (\cdot)
	 | \tbx_{k-1|k-1}
	\Bigg\} \notag \\
	&\leq   \dfrac{(1 + \ol{\Delta}_{\varepsilon})^2 N^3 \ol{h}^2  \ol{\beta}^2 \ol{\vartheta}^2 \bar{c} \bar{r} \bar{p} }{\ul{r}^2}
\end{align}

Now, we let
\begin{align*}
	&\eta_1 = {\ol{\eta}}, \\
	&\eta_2 = \dfrac{N\bar{c}\bar{q}}{ \ul{p}} + \dfrac{(1 + \ol{\Delta}_{\varepsilon})^2 N^3 \ol{h}^2  \ol{\beta}^2 \ol{\vartheta}^2 \bar{c} \bar{r} \bar{p} }{\ul{r}^2}.
\end{align*}

Then, the condition (45) is satisfied. According to Lemma 1, the estimation error $\tbx_{k|k} = \bx_k - \hbx_{k|k}$ of the DUKF-Fc algorithm is exponentially bounded in mean square. The proof is completed.

\section{Numerical Results}

In this section, numerical experiments are conducted to provide a check for the theoretical results. A practical scenario involving single-target tracking using UWSNs is utilized here to validate the effectiveness of DUKF-Fc.

\subsection{Target Tracking using UWSNs}

Tracking maneuvering underwater targets is a primary application of UWSNs with acoustic sensors. In general, a group of underwater sensor nodes (USNs) equipped with battery-powered sensors and local estimators is deployed to follow a non-cooperative underwater target. It is assumed that all the USNs can self-locate and communicate synchronously during the tracking process. For simplicity, USNs are considered homogeneous, sharing identical characteristics of sensors and communication channel. The neighborhood relationships among nodes are established based on communication range, that is, node $j$ is considered an in-neighbor/out-neighbor of node $i$ only if it is within the communication range.

To be specific, we consider an underwater wireless sensor network composed of $N = 20$ USNs in the simulation. The USNs are deployed uniformly in range \qtyproduct{1000 x 1000 x 1500}{\text{m}} with communication range \qty{600}{\text{m}}. The tracking duration is set to be $K = \qty{100}{\text{s}}$ with sampling period $T = \qty{1}{\text{s}}$. The number of simulation runs is set to $M = 1000$. The settings of tracking scenario are summarized in Table~I.

\subsection{Energy Issue}

From equation (3), it is evident that increasing transmission power can enhance transmission reliability and, consequently, state estimation accuracy within a given wireless propagation environment. However, energy conservation is a critical concern for WSNs. As a result, the issue of balancing energy consumption and tracking performance needs to be addressed here.

Suppose that binary frequency shift keying (BFSK) is adopted for the communications between USNs. Following [13], the BER can be denoted as
\begin{align}
	\beta_i( u^{ij}_k g^{ij}_k ) = \Phi \left( \dfrac{u^{ij}_k  g^{ij}_k}{r^i_k k_{\mathrm{B}} \mathrm{T}} \right), \ \forall i \in \cV, j \in \oN_i
\end{align}

\noindent where $r^i_k$ represents the channel bit rate at time instant $k$, $k_{\mathrm{B}}$ is the Boltzmann constant and $\mathrm{T}$ denotes the temperature. The transmission failure probability in (3) thus becomes
\begin{align}
	q^{ij}_k = \left( 1 - \Phi \left( \dfrac{u^{ij}_k  g^{ij}_k}{r^i_k k_{\mathrm{B}} \mathrm{T}} \right) \right)^{l^{ij}_k},
	\label{ppp2}
\end{align}

\noindent and the corresponding transmission energy is given by
\begin{align}
	E^{ij}_k = \dfrac{u^{ij}_k l^{ij}_k}{r^i_k}.
	\label{ppp3}
\end{align}

In the simulation, we assume that parameters including $g^{ij}_k, l^{ij}_k, r^i_k, k_{\mathrm{B}}$ and $\mathrm{T}$ remain constant for $ i \in \cV, k \in \mathbb{N}^+$. In this case, the transmission failure probability can be adjusted by choosing appropriate value of peak power level. Given the parameters' values in Table~II, the relationship between peak power level $u$ and the probability of successful transmission $q$ is depicted in Fig.~1.

\begin{figure}[htb]
	\centering
	\includegraphics[width=3.6in]{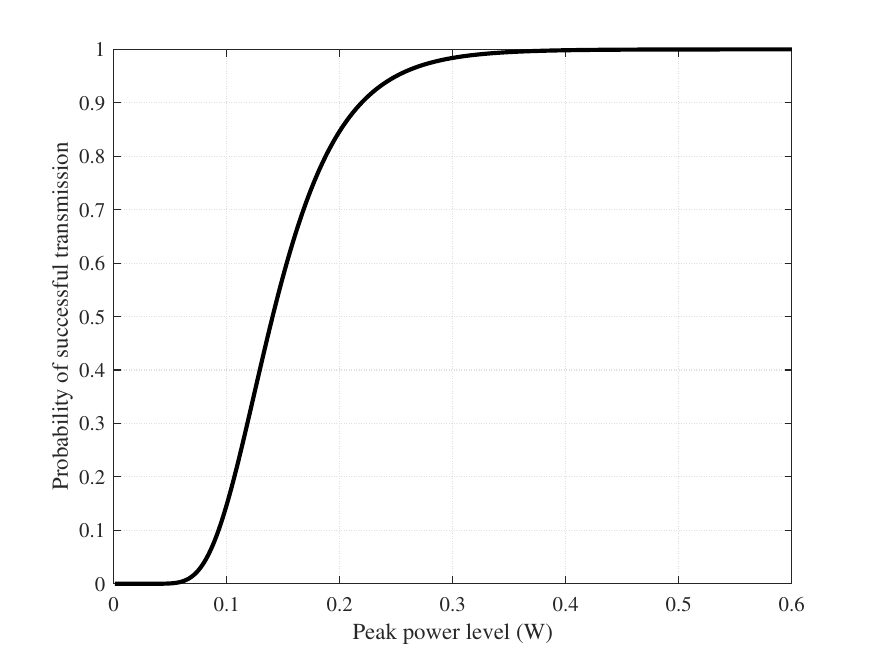}
	\caption{The relationship between peak power level $u$ and transmission failure probability $q$ is illustrated. The values of $g^{ij}_k, l^{ij}_k, r^i_k, k_{\mathrm{B}}, \mathrm{T}$ are given in Table~II. }
	\label{figx-1}
\end{figure}

\subsection{Dynamics}

We consider a target moving at constant velocity (CV) in the $z$-axis and making a constant turn (CT) on the $xy$-plane at a speed of $\omega$ rad/s. The target's state is given by $\bx_k = [x_k, \dot{x}_k , y_k, \dot{y}_k, z_k, \dot{z}_k]^{\tp}$, $(x_k, y_k, z_k)$ being the target's 3D Cartesian coordinates, and $(\dot{x}_k, \dot{y}_k, \dot{z}_k)$ representing the corresponding velocity. The matrices $\bF$ and $\bQ$ are given by:
\begin{align}
	&\bF = \begin{bmatrix}
		\bF_0 & {0}  \\
		{0}  & \bF_1
	\end{bmatrix}, \ \bQ = \begin{bmatrix}
		\bQ_0 & {0} & {0}\\
		{0}  & \bQ_0 & {0} \\
		{0}& {0}& \bQ_0
	\end{bmatrix},
\end{align}

\noindent where
\begin{align*}
	&\bF_0 = \begin{bmatrix}
		1& \sin(\omega T)/\omega&  0& (\cos(\omega T)-1)/\omega \\
		0& \cos(\omega T)& 0&  -\sin(\omega T) \\
		0& (1-\cos(\omega T))/\omega& 1&  \sin(\omega T)/\omega \\
		0& \sin(\omega T)& 0&  \cos(\omega T)
	\end{bmatrix}, \\
	&\bF_1 = \begin{bmatrix}
		1 & T \\
		0 & 1
	\end{bmatrix}, \\
	&\bQ_0 =  \eta^2 \begin{bmatrix}
		T^3/3 & T^2/2 \\
		T^2/2 & T
	\end{bmatrix},
\end{align*}

\noindent while $\eta$ denotes the noise intensity.

The USNs are equipped with sonar arrays, providing the range-only measurements of target. The observation model of node $i$ is
\begin{align}
	\bh^i_k(\bx_k) = \sqrt{(x_k - x^i_k)^2+(y_k - y^i_k)^2+(z_k - z^i_k)^2},
\end{align}

\noindent where $(x^i_k, y^i_k, z^i_k)$ denotes the position of node $i$ at time $k$.

Given initial target state $[0, 10, 0, 3, -1500, 2]^{\tp}$, we set the initial estimation $\hbx^i_{0|0} = [20, -23, 80, 32, -1450, -26]^{\tp}$ and its covariance $\hbP^i_{0|0} = 100 \cdot \bI_6$ for any $i \in \cV$. For conciseness, the settings of other parameters related to tracking are given in Table~III. 

Finally, we use the root mean square error (RMSE) on position and velocity to evaluate the tracking performance, respectively defined as:
\begin{align}
	&\mbox{RMSE}_{p,k} = \sqrt{\dfrac{1}{NM} \sum_{i=1}^{N}\sum_{m=1}^{M}\left( (\tilde{x}^{i, m}_k)^2 + (\tilde{y}^{i, m}_k)^2 + (\tilde{z}^{i, m}_k)^2 \right)}, \\
	&\mbox{RMSE}_{v,k} = \sqrt{\dfrac{1}{NM} \sum_{i=1}^{N}\sum_{m=1}^{M}\left( (\tilde{\dot{x}}^{i, m}_k)^2 + (\tilde{\dot{y}}^{i, m}_k)^2 + (\tilde{\dot{z}}^{i, m}_k)^2 \right)},
\end{align}

\noindent where 
\begin{align*}
	&\tilde{x}^{i, m}_k = x_k - \hat{x}^{i, m}_{k|k}, \ \tilde{\dot{x}}^{i, m}_k = x_k - \hat{\dot{x}}^{i, m}_{k|k}, \\
	&\tilde{y}^{i, m}_k = y_k - \hat{y}^{i, m}_{k|k}, \ \tilde{\dot{y}}^{i, m}_k = y_k - \hat{\dot{y}}^{i, m}_{k|k}, \\
	&\tilde{z}^{i, m}_k = z_k - \hat{z}^{ii, m}_{k|k}, \ \tilde{\dot{z}}^{i, m}_k = z_k - \hat{\dot{z}}^{ii, m}_{k|k},
\end{align*}

\noindent and where $(\hat{x}^{i, m}_{k|k}, {\dot{x}}^{i, m}_k, \hat{y}^{i, m}_{k|k}, {\dot{y}}^{i, m}_k, \hat{z}^{i, m}_{k|k}, {\dot{z}}^{i, m}_k)$ denotes the estimate of target from node $i$ in the $m$th simulation run.

\begin{table}[htb]
	\renewcommand\arraystretch{1.3}
	\centering
	\caption{The settings of UWSNs.}
	\setlength{\abovecaptionskip}{0pt}%    
	\setlength{\belowcaptionskip}{2pt}%
	\label{tab0}
	\begin{tabular}{ l | l  }
		\hline
		{Parameter} & Value \\
		\hline
		Number of USNs, $N$ & \num{20} \\
		Monitored field & \qtyproduct{1000 x 1000 x 1500}{\text{m}} \\
		Communication range & \qty{600}{\text{m}} \\
		Tracking duration, $K$ & \qty{100}{\text{s}}  \\
		Sampling period, $T$ & \qty{1}{\text{s}}  \\
		Number of simulation runs, $M$ & \num{1000}  \\
		\hline
	\end{tabular}
\end{table}

\begin{table}[htb]
	\renewcommand\arraystretch{1.3}
	\centering
	\caption{The parameters related to communication channel.}
	\setlength{\abovecaptionskip}{0pt}%    
	\setlength{\belowcaptionskip}{2pt}%
	\label{tab2}
	\begin{tabular}{ l | l  }
		\hline
		{Parameter} & Value \\
		\hline
		Channel power gain, $g^{ij}_k$ & \qty{-150}{\text{dB}} \\
		Number of bits, $l^{ij}_k$ & \qty{1000}{\text{bits}} \\
		Channel bit rate, $r^i_k$ & \qty{6000}{\text{bits}\cdot\text{s}^{-1}}  \\
		Boltzmann constant, $k_{\mathrm{B}}$ & \qty{1.38e-23}{\text{J}\cdot\text{K}^{-1}}  \\
		Temperature, ${\mathrm{T}}$ & \qty{280}{\text{K}}   \\
		\hline
	\end{tabular}
\end{table}

\begin{table}[htb]
	\renewcommand\arraystretch{1.3}
	\centering
	\caption{The Parameters related to tracking.}
	\setlength{\abovecaptionskip}{0pt}%    
	\setlength{\belowcaptionskip}{2pt}%
	\label{tab1}
	\begin{tabular}{ l | l  }
		\hline
		{Parameter} & Value \\
		\hline
		Turning speed, $\omega$ & \qty{0.52}{\text{rad}\cdot \text{s}^{-1}}  \\
		Intensity of process noise, $\eta^2$  & $5$   \\
		Covariance of measurement noise, $\bR^i_{v}$  & $10\sqrt{i} \cdot \bI_1$   \\
		Covariance of additional noise, $\bR^i_{n}$  & $\sqrt{i} \cdot \bI_1$   \\
		Covariance of fading coefficient, $\sigma^{i}_{\vartheta}$ & $0.5$ \\
		Covariance of relative estimation error, $\sigma^{i}_{\varepsilon}$ & $\sqrt{i} \cdot \bI_1$ \\
		Bound of relative error, $\Delta^i_{\varepsilon}$ & $10\%$ \\
		\hline
	\end{tabular}
\end{table}

\subsection{Results}

Fig.~2 illustrates the real target trajectory and the estimated trajectory using DUKF-Fc with peak power levels $u^{ij}_k$ in range   $[\qty{118}{\text{mW}}, \qty{168}{\text{mW}}] $ ($q^{ij}_k \in [0.3, 0.7]$), $\forall i, j, k.$ The estimated trajectory is obtained by averaging a total of 20 estimates from USNs. It is demonstrated that the proposed DUKF-Fc can effectively estimate the target trajectory.

\begin{figure}[htb!]
	\centering
	\includegraphics[width=3.45in]{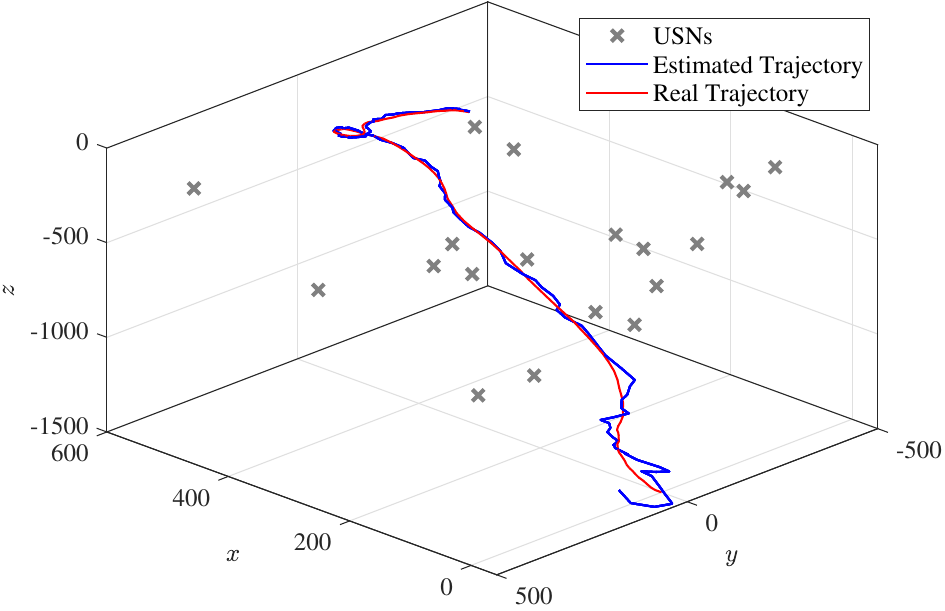}
	\caption{A snapshot of estimating the target trajectory using the proposed DUKF-Fc with $u^{ij}_k$ in range  $[\qty{118}{\text{mW}}, \qty{168}{\text{mW}}] $ ($q^{ij}_k \in [0.3, 0.7]$), $\forall i \in \cV, j \in \oN, k \in \mathbb{N}^+.$ The estimated trajectory is obtained by averaging a total of 20 estimates from USNs.}
	\label{figx1}
\end{figure}

Fig.~3 and Fig.~4 compare position and velocity RMSEs of the proposed DUKF-Fc with those of other methods. The peak power levels $u^{ij}_k$ are uniformly chosen from range $[\qty{135}{\text{mW}}, \qty{147}{\text{mW}}] $ ($q^{ij}_k \in [0.45, 0.55]$), $\forall i, j, k.$ Here, we denote the case using exact values of fading coefficients as DUKF-eFc and the case without considering the effects of measurement fluctuation as DUKF-nFc. It can be noticed that the RMSEs of the proposed DUKF-Fc are bounded in mean-square. Also, the performance of DUKF-Fc significantly outperforms that of DUKF-nFc and is comparable to that of DUKF-eFc. However, it should be mentioned that the convergence speed of DUKF-Fc is found to be slower than DUKF-eFc due to the channel estimation error. 

\begin{figure}[htbp!]
	\centering
	\includegraphics[width=3.6in]{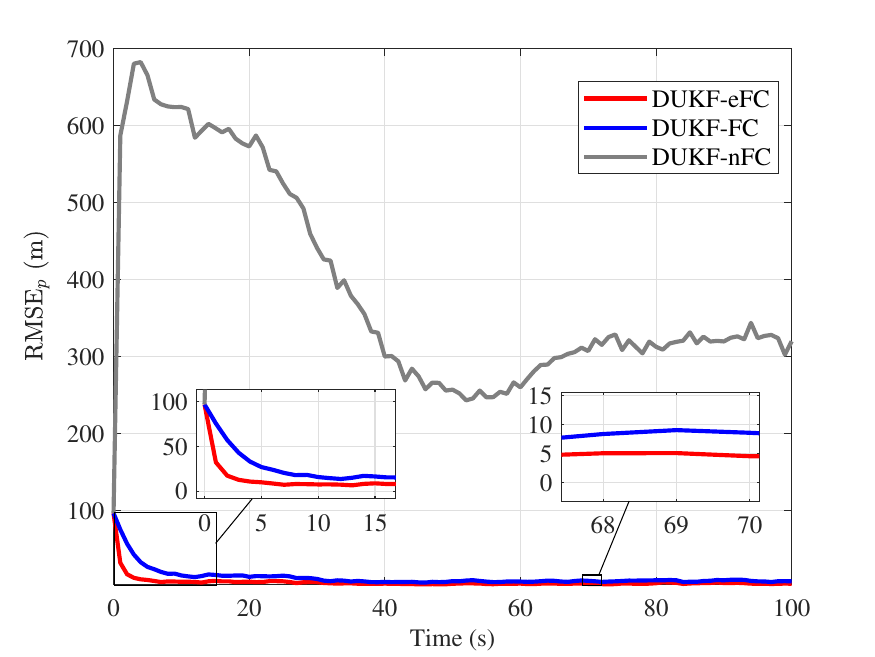}
	\caption{The RMSE of position using different methods with peak power level $u^{ij}_k \in [\qty{118}{\text{mW}}, \qty{168}{\text{mW}}] $ ($q^{ij}_k \in [0.45, 0.55]$), $\forall i \in \cV, j \in \oN, k \in \mathbb{N}^+.$ }
	\label{figx4}
\end{figure}

\begin{figure}[htbp!]
	\centering
	\includegraphics[width=3.6in]{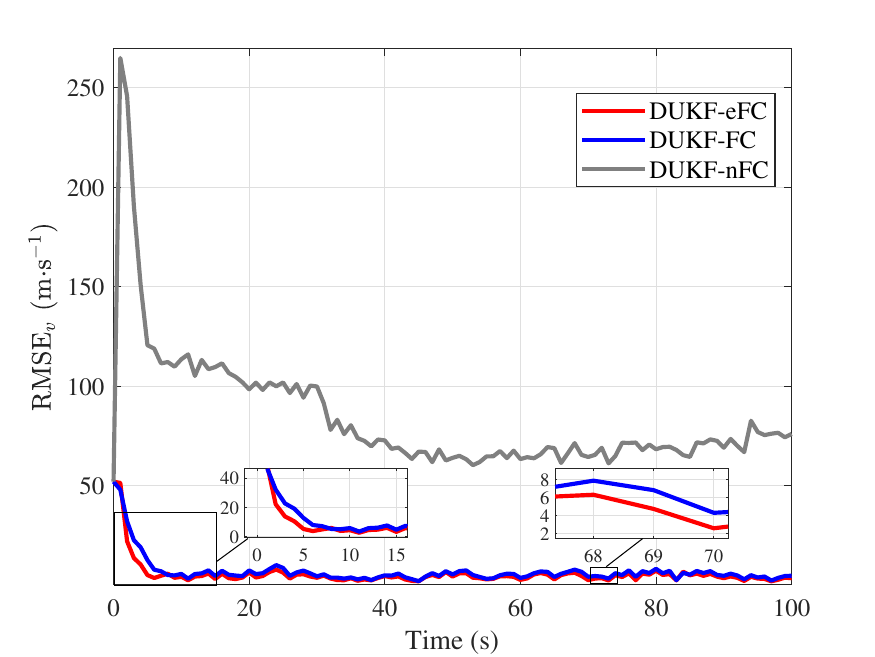}
	\caption{The RMSE of velocity using different methods with peak power level $u^{ij}_k \in [\qty{118}{\text{mW}}, \qty{168}{\text{mW}}] $ ($q^{ij}_k \in [0.45, 0.55]$), $\forall i \in \cV, j \in \oN, k \in \mathbb{N}^+.$ }
	\label{figx5}
\end{figure}

Fig.~5 and Fig.~6 show the RMSEs of the proposed DUKF-Fc with different values of power level. The peak power levels of USNs are set to be homogeneous and thus, the marks $(i,j,k)$ on $u^{ij}_k$ and $q^{ij}_k$ are omitted herein. It is obvious that the RMSEs decrease for higher peak power level. Besides,  the performance loss degrades when the transmission failure probability approaches $1$. Detailed comparisons of RMSEs are presented in Table~IV and Table~V, which confirm the aforementioned observations.

\begin{table*}[htbp!]
	\setlength{\tabcolsep}{10pt}
	\renewcommand\arraystretch{1.20}
	\centering
	\caption{Accuracy of position estimation.}
	\setlength{\abovecaptionskip}{0pt}%    
	\setlength{\belowcaptionskip}{2pt}%
	\label{tabx1}
		\begin{tabular}{ l | ccccc }
			\hline
			RMSE$_{p}$ (m) & $\begin{aligned}[c]
				&u = \qty{93}{\text{mW}}  \\[-0.1cm] 
				&\ (q = 0.1)
			\end{aligned}$  & $\begin{aligned}[c]
			&u = \qty{118}{\text{mW}}  \\[-0.1cm] 
			&\ (q = 0.3)
		\end{aligned}$ & $\begin{aligned}[c]
		&u = \qty{140}{\text{mW}}  \\[-0.1cm] 
		&\ (q = 0.5)
	\end{aligned}$ & $\begin{aligned}[c]
	&u = \qty{168}{\text{mW}}  \\[-0.1cm] 
	&\ (q = 0.7)
\end{aligned}$ & $\begin{aligned}[c]
&u \geq \qty{400}{\text{mW}}  \\[-0.1cm] 
&\ (q \approx 1.0)
\end{aligned}$ \\
			\hline
			$\mbox{DUKF-CF}$ & $26.744$  & $14.913$ & $12.824$ & $11.760$ & $11.273$ \\
			$\mbox{DUKF-nCF}$ & $449.023$  & $394.655$ & $387.262$ & $376.385$ & $343.762$ \\
			$\mbox{DUKF-eCF}$ & $17.558$  & $8.249$ & $7.030$ & $6.577$ & $6.423$ \\
			\hline
		\end{tabular}
\end{table*}

\begin{table*}[htbp!]
		\setlength{\tabcolsep}{10pt}
		\renewcommand\arraystretch{1.20}
		\centering
		\caption{Accuracy of velocity estimation.}
		\setlength{\abovecaptionskip}{0pt}%    
		\setlength{\belowcaptionskip}{2pt}%
		\label{tabx2}
		\begin{tabular}{ l | ccccc }
			\hline
			RMSE$_{v}$ (m$\cdot$s$^{-1}$) & $\begin{aligned}[c]
				&u = \qty{93}{\text{mW}}  \\[-0.1cm] 
				&\ (q = 0.1)
			\end{aligned}$  & $\begin{aligned}[c]
				&u = \qty{118}{\text{mW}}  \\[-0.1cm] 
				&\ (q = 0.3)
			\end{aligned}$ & $\begin{aligned}[c]
				&u = \qty{140}{\text{mW}}  \\[-0.1cm] 
				&\ (q = 0.5)
			\end{aligned}$ & $\begin{aligned}[c]
				&u = \qty{168}{\text{mW}}  \\[-0.1cm] 
				&\ (q = 0.7)
			\end{aligned}$ & $\begin{aligned}[c]
				&u \geq \qty{400}{\text{mW}}  \\[-0.1cm] 
				&\ (q \approx 1.0)
			\end{aligned}$ \\
			\hline
			$\mbox{DUKF-FC}$ & $9.100$  & $7.447$ & $6.693$ & $6.685$ & $6.681$ \\
			$\mbox{DUKF-nFC}$ & $107.375$  & $98.046$ & $95.553$ & $89.775$ & $83.176$ \\
			$\mbox{DUKF-eFC}$ & $7.577$  & $6.018$ & $5.594$ & $5.389$ & $5.247$ \\
			\hline
		\end{tabular}
\end{table*}

\begin{table*}[htbp!]
	\setlength{\tabcolsep}{8pt}
	\renewcommand\arraystretch{1.20}
	\centering
	\caption{The change rates of RMSEs and energy consumption.}
	\setlength{\abovecaptionskip}{0pt}%    
	\setlength{\belowcaptionskip}{2pt}%
	\label{tabx3}
	
	\begin{tabular}{c | c c c c c c c }
		\hline
		$u$ (mW)&  $q$ & RMSE$_{p}$ (m)& Change rate & RMSE$_{v}$ (m$\cdot$s$^{-1}$)& Change rate & Energy cost (J$\cdot$s$^{-1}$) & Change rate \\
		\hline
		$93$ &$0.1$ & $26.744$ & $+137.20\%$ & $9.100$ & $+36.21\%$& $2.945$ & $-76.16\%$ \\
		\hline
		$118$ &$0.3$ & $14.913$ & $+32.29\%$ & $7.447$ & $+11.47\%$& $3.737$ & $-69.75\%$ \\
		\hline
		$140$ &$0.5$ & $12.824$ & $+13.76\%$ & $6.693$ & $+0.18\%$& $4.323$ & $-65.00\%$\\
		\hline
		$168$ &$0.7$ & $11.760$  & $+4.32\%$ & $6.685$ & $+0.06\%$& $5.126$ & $-58.50\%$\\
		\hline
		$400$ &$\approx 1.0$ & $11.273$ & $0\%$ & $6.681$ & $0\%$ & $12.352$ & $0\%$\\
		\hline
	\end{tabular}
\end{table*}

\begin{figure}[htbp!]
	\centering
	\includegraphics[width=3.6in]{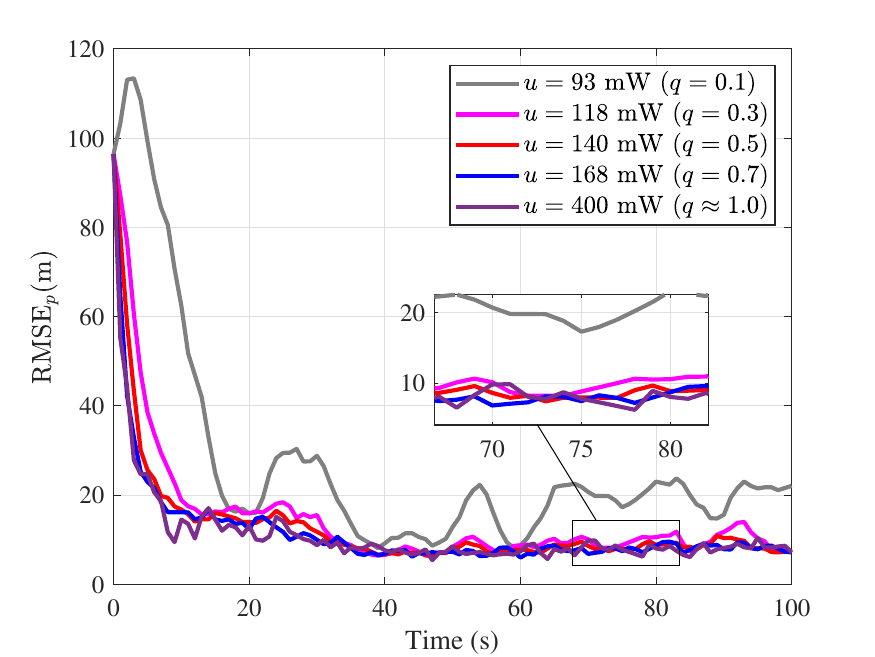}
	\caption{The RMSE of position under different values of $q$.}
	\label{figx2}
\end{figure}

\begin{figure}[htbp!]
	\centering
	\includegraphics[width=3.6in]{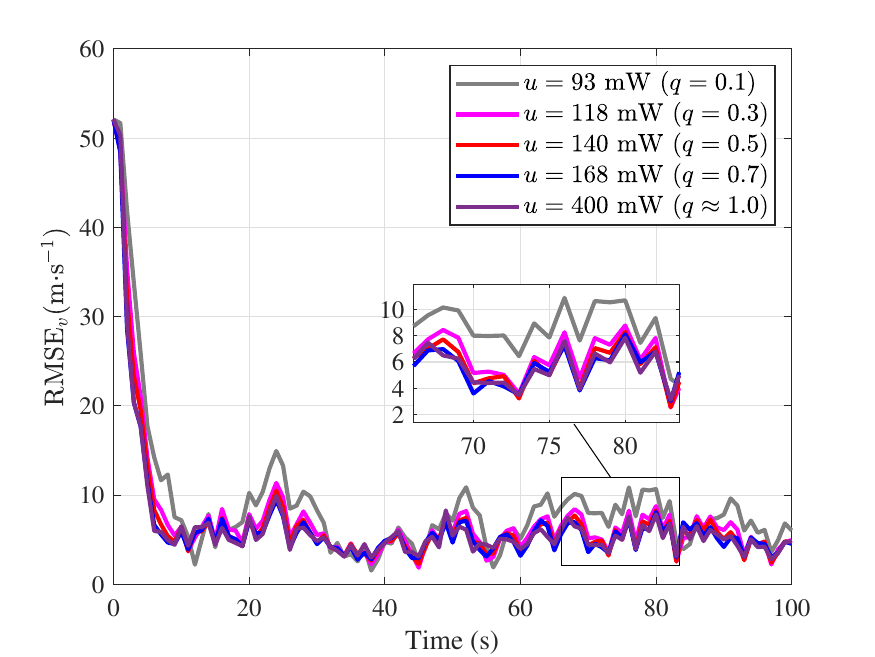}
	\caption{The RMSE of velocity under different values of $q$.}
	\label{figx3}
\end{figure}

From Fig.~5, Fig.~6 and the data in Table~IV and Table~V, it can be noticed that there is a trade-off between the tracking performance and the energy cost. Therefore, to verify the energy-efficiency of the DUKF-Fc algorithm, we compute the change rates of RMSEs and the energy consumption compared to the error-free case ($u = \qty{400}{\text{mW}}$ ($q \approx 1.0$) ). The results are presented in Table~VI.

From Table~VI,  it is seen that decreasing the peak power level to certain degrees results in slight performance degradation, however, significant reduction in the energy consumption with respect to the error-free case. For example, the proposed DUKF-Fc algorithm gives 58.50\% energy saving at the expense of only 4.32\% and 0.06\% performance loss in terms of RMSEs (with reference to $u = \qty{400}{\text{mW}}$ ($q \approx 1.0$) ) for $u = \qty{168}{\text{mW}}$ ($q \approx 0.7$).

\section{Conclusion}

In this paper, we propose a modified DUKF under channel fading, called DUKF-Fc, for distributed target tracking. The filter is built upon the framework of standard UKF and the effects of channel fading including measurement fluctuation and transmission failure are considered and compensated in the filter design. In view of the fact that the fading coefficients are a priori unknown and can only be estimated in real applications, the channel estimation error is also taken into account. Further, we perform theoretical analysis and prove that the estimation error is exponentially bounded in mean square for the distributed nonlinear filter. Finally, the DUKF-Fc algorithm is applied to underwater target tracking using UWSNs, serving as a means to validate its efficacy. The simulation results demonstrate that a desired balance between filtering performance and energy consumption can be achieved by properly adjusting the peak power level of transmitters, but more detailed analysis should be done in the future to further investigate the scheduling of energy resources.

%\bibliographystyle{IEEEtran}
%\bibliography{IEEEabrv, bare_jrnl_new_sample4}


\vfill

\end{document}